\documentclass[twoside]{ae100prg}
\usepackage{graphicx}
\usepackage[breaklinks]{hyperref}
\usepackage{booktabs}

\begin{document}
\title{Shape Dynamics}


\author{Tim A. Koslowski$^1$}

\address{$^1$ Perimeter Institute for Theoretical Physics, 31 Caroline St. N, Waterloo, Ontario, N2L 2Y5, Canada\\New address: Department of Mathematics and Statistics, University of New Brunswick, Fredericton, New Brunswick E3B 5A3, Canada}

\email{$^1$ t.a.koslowski@gmail.com}

\begin{abstract}
  Barbour's formulation of Mach's principle requires a theory of gravity to implement local relativity of clocks, local relativity of rods and spatial covariance. It turns out that relativity of clocks and rods are mutually exclusive. General Relativity implements local relativity of clocks and spatial covariance, but not local relativity of rods. It is the purpose of this contribution to show how Shape Dynamics, a theory that is locally equivalent to General Relativity, implements local relativity of rods and spatial covariance and how a BRST formulation, which I call Doubly General Relativity, implements all of Barbour's principles.
\end{abstract}

\section{Introduction}

A reflection on Mach's principle led Barbour to postulate that rods and spatial frames of reference should be locally determined by a procedure that he calls ``best matching,'' while clocks should be locally determined by what he calls ``objective change'' (for more see \cite{Barbour:2012XX}). More concretely, Barbour's principles postulate local time reparametrization invariance, local spatial conformal invariance and spatial covariance. The best matching algorithm for spatial covaraince and local spatial conformal invariance turns out to be equivalent to the imposition of linear diffeomorphism and conformal constraints
\begin{equation}
 H(\xi)=\int_\Sigma d^3x\, \pi^{ab}(\mathcal L_\xi g)_{ab},\,\,\,\,C(\rho)=\int_\Sigma d^3x\, \rho\, \pi,
\end{equation}
where we use a compact Cauchy surface $\Sigma$ without boundary with Riemannian metric $g_{ab}$ and metric momentum density $\pi^{ab}$ with trace $\pi$. The vector field $\xi$ and the scalar field $\rho$ are Lagrange multipliers. A more involved procedure, which I will not explain here, leads to the implementation of local time reparametrization invariance through quadratic Hamilton constraints
\begin{equation}
 \hat S(N)=\int_\Sigma d^3x\,N\,\left(\pi^{ab}F_{abcd}\pi^{cd}-V\right),
\end{equation}
where $F_{abcd}(x)$ and $V(x)$ are constructed from $g_{ab}(x)$ and its derivatives at $x$ and $N$ denotes a Lagrange multiplier. There is no reason for $\hat S$ to have homogeneous conformal weight, so a system containing the constraints $\hat S(N)$ and $C(\rho)$ will not be first class except for very special choices of $F_{abcd},V$. This means that time reparametrization symmetry and spatial conformal symmetry generically exclude one another. An interesting situation occurs when we choose the $F_{abcd},V$ to reproduce the Hamilton constraints of General Relativity
\begin{equation}
 S(N)=\int_\Sigma d^3x\,N\,\left(\frac{\pi^{ab}(g_{ac}g_{bd}-\frac 1 2g_{ab}g_{cd})\pi^{cd}}{\sqrt{|g|}}-(R-2\Lambda)\sqrt{|g|}\right),
\end{equation}
where the constraint system $S(N),Q(\rho)$ is completely second class, while the constraint system $S(N),H(\xi)$ is first class as is the constraint system $Q(\rho),H(\xi)$. We will shortly see that this is the reason why Shape Dynamics and Doubly General Relativity can be constructed \cite{Gomes:2010fh}.

\section{Symmetry Trading}

Gauge theories describe a physical system using redundant degrees of freedom. The physical degrees of freedom are identified with orbits of the action of the gauge group. This redundant description is very useful in field theory, because it is often the only local description of a given system. The canonical description of gauge theories (see e.g. \cite{Henneaux:1992ig}) is provided by a regular irreducible set of first class constraints $\{\chi_\alpha\}_{\alpha\in\mathcal A}$, whose elements $\chi_\alpha$ are smooth functions on a phase space $\Gamma$ with Poisson bracket $\{.,.\}$. First class means that the constraint surface $\mathcal C=\{x\in \Gamma:\chi_\alpha(x)=0,\,\forall \alpha \in \mathcal A\}$ is foliated into gauge orbits, whose infinitesimal generators are the Hamilton vector fields $v_\alpha:f\mapsto \{\chi_\alpha,f\}$. For simplicity we will assume that the system is generally covariant, so it has vanishing on-shell Hamiltonian. Observables $[O]$ of the system are equivalence classes of 
smooth gauge-invariant functions $O$ on $\Gamma$, where two functions are equivalent 
if they coincide on $\mathcal C$ and where gauge invariance means that $O$ is constant along gauge orbits. This means that an observable is completely determined by determining its dependence on a reduced phase space $\Gamma_{red}$, which contains one and only one point out of each gauge orbit. There is no unique choice of $\Gamma_{red}$ and the simplest description is through a regular irreducible set of gauge fixing conditions $\{\sigma^\alpha\}_{\alpha \in \mathcal A}$ such that a proper reduced phase is defined through $\Gamma_{red}=\{x\in \Gamma:\chi_\alpha(x)=0=\sigma^\alpha(x),\,\forall \alpha \in \mathcal A\}$. The observable algebra can then be identified with the Dirac algebra on reduced phase space, where the Dirac bracket takes the form
\begin{equation}
 \{f,g\}_D:=\{f,g\}-\left(\{f,\chi_\alpha\}M^\alpha_\beta\{\sigma^\beta,g\}-\{f,\sigma^\beta\}M^\alpha_\beta\{\chi_\alpha,g\}\right),
\end{equation}
where $M$ denotes the inverse of the linear operator $\{\chi,\sigma\}$.

The condition that $\Gamma_{red}$ contains one and only one point out of each gauge orbit imposes important restrictions on the gauge fixing conditions, but the set of gauge fixing conditions is \emph{not} required to be first class. A very interesting situation arises when the set of gauge fixing conditions is itself first class: In this case one can switch the role of gauge fixing conditions and constraints and describe the same observable algebra, and thus the same physical system, with the gauge theory $A=(\Gamma,\{.,.\},\{\chi_\alpha\}_{\alpha\in \mathcal A})$ or with the gauge theory $B=(\Gamma,\{.,.\},\{\sigma^\alpha\}_{\alpha\in \mathcal A})$. The manifest equivalence of the two theories is established by gauge fixing theory $A$ with the gauge fixing set $\{\sigma^\alpha\}_{\alpha \in \mathcal A}$ and gauge theory $B$ with the gauge fixing set $\{\chi_\alpha\}_{\alpha \in \mathcal A}$. One thus trades one gauge symmetry for another. 

\begin{figure}
  \begin{center}
  \includegraphics[width=9cm]{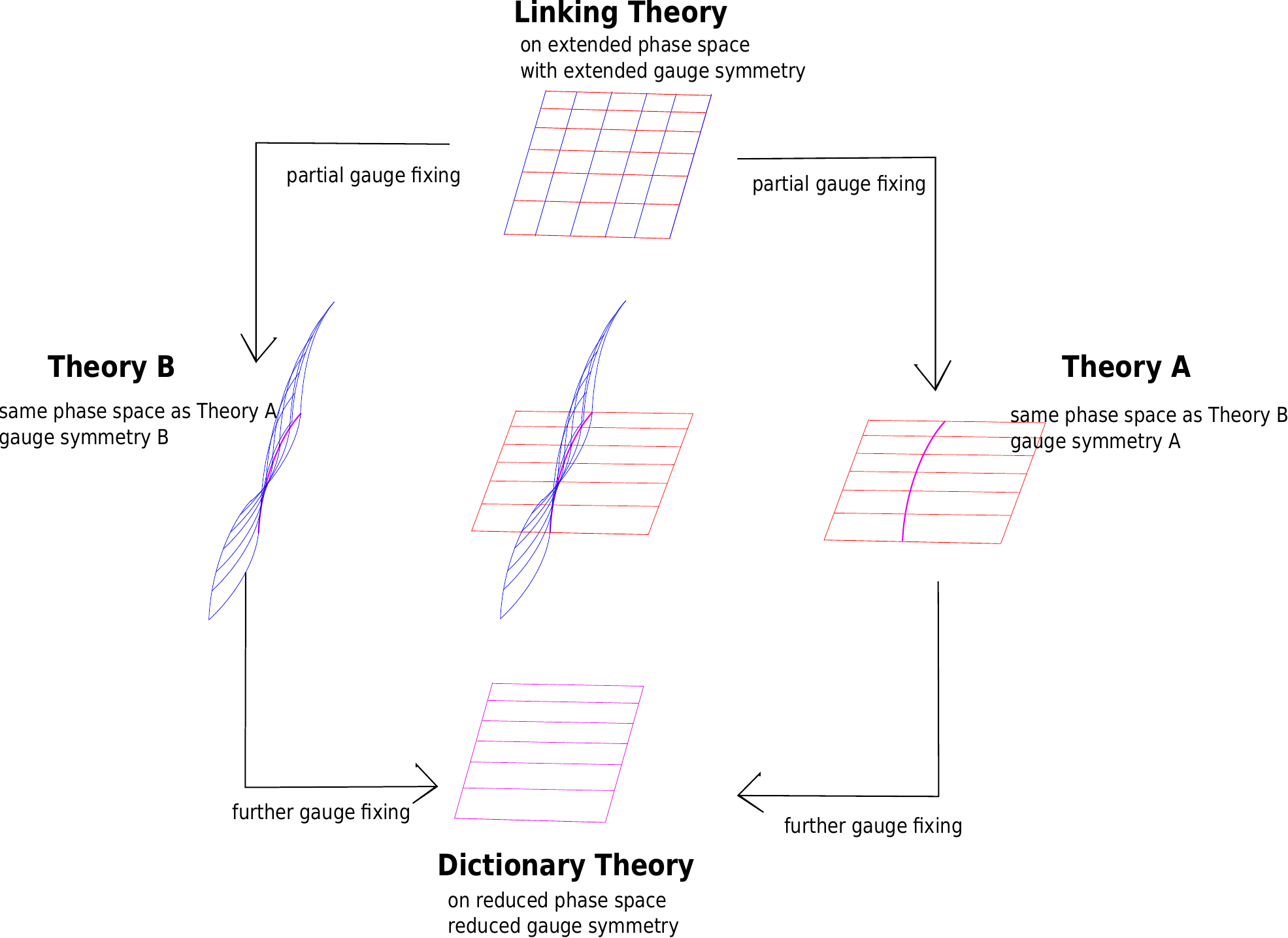}
  \end{center}
  \caption{\label{figure:LT} Construction of equivalent gauge theories from a linking theory.}
\end{figure}

A very useful tool for the construction of equivalent gauge theories is a \emph{linking theory}, see \cite{Gomes:2011zi} and figure \ref{figure:LT}. 
Let us start with a phase space extension and denote the configuration variables of the extension by $\phi_\alpha$ and their canonically conjugate momenta by $\pi^\beta$ and local Darboux coordinates on the original phase space $\Gamma$ by $(q_i,p^j)$. A linking theory on extended phase space is a set of regular irreducible first class constraints that can be split into three subsets: The set $\{\chi^1_\alpha\}_{\alpha \in \mathcal A}$ can be weakly\footnote{``Weakly'' means on the constraint surface.} solved for $\phi_\alpha$, the set $\{\chi^\alpha_2\}_{\alpha \in\mathcal A}$ can be weakly solved for $\pi^\alpha$ and the set $\{\chi^3_\mu\}_{\mu \in \mathcal M}$ is weakly independent of the phase space extension. In this case, we can simplify the discussion by noticing that the three constraint sets are equivalent to the sets
\begin{equation}
 \{\phi_\alpha-\phi_o^\alpha(q,p)\}_{\alpha \in \mathcal A},\,\,\,\{\pi^\alpha-\pi^o_\alpha(q,p)\}_{\alpha \in \mathcal A}\,\textrm{and}\,\,\{\tilde{\chi}^3_\mu(q,p)\}_{\mu \in \mathcal M}.
\end{equation}
There are two sets of natural gauge fixing conditions $\{\phi_\alpha\}_{\alpha \in \mathcal A}$ and $\{\pi^\alpha\}_{\alpha \in \mathcal A}$. Imposing the gauge $\phi_\alpha=0$ fixes the constraints $\pi^\alpha-\pi^o_\alpha(q,p)$ and leads to the phase space reduction $(\phi_\alpha,\pi^\beta) \to (0,\pi^o_\beta(q,p))$, so the reduced phase space is $\Gamma$. Moreover, the Dirac bracket associated with this phase space reduction coincides with the Poisson bracket on $\Gamma$. The result of the phase space reduction is the gauge theory $B=(\Gamma,\{.,.\},\{\pi^o_\alpha\}_{\alpha \in \mathcal A}\cup\{\tilde\chi^3_\mu\}_{\mu \in \mathcal M})$.

Similarly, imposing $\pi^\alpha=0$ yields a phase space reduction $(\phi_\alpha,\pi^\beta)\to(\phi_o^\alpha(q,p),0)$ and the resulting gauge theory is $A=(\Gamma,\{.,.\},\{\phi^o_\alpha\}_{\alpha \in \mathcal A}\cup\{\tilde\chi^3_\mu\}_{\mu \in \mathcal M})$. The gauge theories $A$ and $B$ obviously describe the same physical system. It turns out that we would have obtained the same result even if had we not solved the first two subsets of constraints for the phase space extension.  

\section{Shape Dynamics}

Let us now extend the phase space of General Relativity by a conformal factor $\phi$ and its conjugate momentum density $\pi_\phi$. The linking theory between General Relativity on a compact manifold $\Sigma$ without boundary and Shape Dynamics can be obtained by canonical best matching General Relativity in the ADM formulation with respect to conformal transformations that do not change the total spatial volume. This yields the following set of constraints
\begin{equation}
 \begin{array}{l}
   TS(N)=\int_\Sigma d^3xN\left(\frac{\sigma^a_b\sigma^b_a}{\sqrt{|g|}}e^{-6\hat\phi}+(2\Lambda-\frac 1 6 \langle \pi\rangle^2) \sqrt{|g|}e^{6\hat\phi}-R(e^{6\hat\phi}g)\sqrt{|g|}e^{2\hat\phi}+a\right)\\
   Q(\rho)=\int_\Sigma d^3x\,\rho\,\left(\pi_\phi-4\pi+4\langle\pi\rangle\sqrt{|g|}\right)\\
   H(\xi)=\int_\Sigma d^3x \left(\pi^{ab}(\mathcal L_\xi g)_{ab}+\pi_\phi\mathcal L_\xi \phi\right),
 \end{array}
\end{equation}
where $\hat \phi:=\phi-\frac 1 6 \ln\langle e^{6\phi}\rangle$, $\sigma^a_b=\pi^{ac}g_{cb}-\frac 1 3 \pi \delta^a_b$ and where triangle brackets denote the mean w.r.t. $\sqrt{|g|}$ and where the term $a$ vanishes when $\pi=\langle\pi\rangle\sqrt{|g|}$. Imposing the gauge fixing condition $\phi=0$ results in a phase space reduction $(\phi,\pi_\phi)\to(0,4(\pi-\langle\pi\rangle))$ and reduces the system back to the ADM formulation of General Relativity. Imposing the gauge fixing condition $\pi_\phi=0$ gauge fixes all $TS(N)$ except for one. The simplest way to see this is to observe that $TS=0$ is equivalent to requiring that $\phi$ solves the Lichnerowicz-York equation
\begin{equation}
 8\Delta_g \Omega=\left(\frac 1 6 \langle \pi\rangle^2-2\Lambda\right)\,\Omega^5+R\,\Omega-\frac{\sigma^a_b\sigma^b_a}{|g|}\,\Omega^{-7}
\end{equation}
for $\Omega=e^\phi$ but with the reducibility condition that the conformal factor is volume preserving $\int_\Sigma d^3x \sqrt{|g|}\left(1-e^{6\phi}\right)=0$. The left-over constraint is thus equivalent to the constraint 
\begin{equation}
 \int_{\Sigma} d^3x \sqrt{|g|}\left(1-e^{6\phi_o[g,\pi]}\right)=0,
\end{equation}
where $\phi_o[g,p;x)$ denotes the positive solution to the Lichnerowicz-York equation, which is known to uniquely exist on physical phase space \cite{OMurchadha:1973XX}. The phase space reduction thus yields the constraint system
\begin{equation}
 \begin{array}{rcl}
   H_{SD}&=&\int_\Sigma d^3x \sqrt{|g|}(1-e^{6\phi_o[g,\pi]})\\
   \hat C(\rho)&=&\int_\Sigma d^3x\,\rho\,\left(\pi-\langle\pi\rangle\sqrt{|g|}\right)\\
   H(\xi)&=&\int_\Sigma d^3x \pi^{ab}(\mathcal L_\xi g)_{ab}.
 \end{array}
\end{equation}
This is \emph{not exactly} Shape Dynamics because the total conformal transformations generated by $\hat C(\rho)$ preserve the total spatial volume. One can however obtain a \emph{true} theory of Shape Dynamics by observing that the only nonlinear constraint $H_{SD}$ has the form of a reparametrization constraint $p_t-H(t)\approx 0$ of parametrized dynamics. Thus, after identifying the total volume $V$ with the momentum conjugate to York time $\tau=\frac{3}{2}\langle\pi\rangle$ and \emph{deparametrizing} the theory one obtains the physical Hamiltonian
\begin{equation}
 H_{phys}=\int_\Sigma \sqrt{|g|} e^{6\phi[g,\pi]}.
\end{equation}
The $\pi(x)$ is constrained to $\langle \pi\rangle\sqrt{|g|(x)}$ and the conformal factor of the metric is pure gauge, except for the total volume. The physical phase space is thus coordinatized by the conformal metric $\rho_{ab}=|g|^{-1/3}g_{ab}$, trace free momenta $\sigma^a_b$ and the pair $V,\langle \pi\rangle$. The only physical phase space coordinate that is affected differently by volume preserving conformal transformations as opposed to unrestricted conformal transformations is $V$. But after reinterpreting $\frac 3 2 \langle\pi\rangle,V$ as time and its momentum there is \emph{no} difference between the physical phase space volume preserving and unrestricted conformal transformations. The Shape Dynamics Hamiltonian $H_{phys}$ comes with the constraints
\begin{equation}
  C(\rho)=\int_\Sigma d^3x\,\pi,\,\,\,H(\xi)=\int_\Sigma d^3x\,\pi^{ab}(\mathcal L_\xi g)_{ab}.
\end{equation}
The dictionary between Shape Dynamics and General Relativity is established as follows: Given a solution $\rho_{ab}(\tau),\sigma^a_b(\tau)$ to the Shape Dynamics equations of motion, one finds that this is also a solution to the equations of motion of General Relativity in constant mean curvature (CMC) gauge. 

\section{Doubly General Relativity}

An important difficulty of Shape Dynamics is that the physical Hamiltonian is non-local. One can improve this situation by working with the BRST formalism. One obtains the BRST formalism by adjoining to each of the regular irreducible first class constraints $\chi_\alpha$ a ghost $\eta^\alpha$ and canonically conjugate ghost momentum $P_\alpha$ with opposite statistics. A result from cohomological perturbation theory then shows that the first class property of the constraints allows one to construct a nilpotent ghost number one BRST generator $\Omega=\eta^\alpha\chi_\alpha+\mathcal O(\eta^2)$, which provides a resolution of the observable algebra. This means that $\{\Omega,.\}$ is a differential whose cohomology at ghost number zero is precisely the classical observable algebra.

The prerequisite for symmetry trading to work was that there were two first class surfaces that gauge fixed one another. One can thus construct two nilpotent BRST generators: A ghost number $+1$ generator $\Omega$ from the first set of first class constraints and a ghost number $-1$ generator $\Psi$ from the second set of first class constraints. For generally covariant theories, i.e. theories with vanishing on-shell Hamiltonian, one finds that the BRST gauge-fixed Hamiltonian can be written as $H_{BRST}=\{\Omega,\Psi\}$. The Jacobi identity and nilpotency of the generators then imply that $H_{BRST}$ is annihilated by two BRST transformations: those generated by $\Omega$ and those generated by $\Psi$. One can thus not only relate the observables of the two theories with one another ($\Omega$ provides a resolution for the first and $\Psi$ for the second), but one also sees that the BRST gauge-fixed actions of the two theories can be chosen to coincide.

It goes beyond the scope of this contribution to discuss a detailed application of this construction for the duality between General Relativity and Shape Dynamics (for details see \cite{Gomes:2012hh}), so I will only illustrate what the two BRST charges can be chosen to be:
\begin{equation}
 \begin{array}{rcl}
   \Omega&=&\int_\Sigma d^3x \left( S(\eta) +H^a(\eta_a)+\mathcal O(\eta^2)\right)\\
   \Psi&=&\int_\Sigma d^3x\left( P \frac{\pi}{\sqrt{|g|}}+\frac{H^a(g_{ab}P^b)}{\sqrt{|g|}}+\mathcal O(P^2) \right),
  \end{array}
\end{equation}
where the higher orders in ghosts are chosen such that the two generators are nilpotent. It is evident that the construction will lead to a canonical action that is left invariant by ADM- and Shape Dynamics BRST transformations, hence the name Doubly General Relativity. I conclude this short description with the warning that the ghost-number zero term of this Hamiltonian is {\emph{not}} the CMC gauge-fixing of ADM.

\section{Interpretation}

Although local relativity of clocks and local relativity of rods seem to be incompatible as gauge symmetries, they are reconcilable. This reconciliation can be seen in the canonical formalism, where it appears that gravitational dynamics is equivalently described either by the ADM system or by the Shape Dynamics system. This means that both theories have the same solutions and make the same predictions for all observables. It can also be seen in the BRST formalism, where not only the predictions for observables coincide, but where it turns out that  the gauge fixed gravity action has two BRST invariances, one corresponding to the on-shell spacetime symmetries of the ADM description of gravity, the other to corresponding to the conformal symmetries of the Shape Dynamics description of gravity.

I want to conclude by considering the ideas behind the construction of Shape Dynamics. The construction of Shape Dynamics requires two steps: symmetry trading and the identification of a parametrized dynamical system. There is a lot of literature on the second step (e.g. Kucha\v{r}'s perennials), but the idea behind symmetry trading seems to have not been discussed in the literature although the idea itself is obvious.

Gauge theories are formulated with redundant degrees of freedom and gauge invariance to have a local\footnote{Note that the Shape Dynamics Hamiltonian, although nonlocal, can be described locally through the linking theory.} (and thus comprehensible) field theory. This is a special instance of the more general fact that a comprehensible description of the real world often requires auxiliary concepts that are not truly part of reality. Which auxiliary concept is chosen is not unique; in general there is an infinite number of internally consistent descriptions. But although all descriptions are required to accurately describe the real world and to be internally consistent, it can happen that two descriptions are mutually exclusive because the auxiliary concepts are incompatible. We see this in the duality between General Relativity and Shape Dynamics: General Relativity teaches that gravity is spacetime geometry and not a conformal theory, while Shape Dynamics teaches that gravity is a conformal theory without 
spacetime.

This can be very disturbing and leads to the question: ``How can we discriminate between better and worse descriptions?'' Provided two descriptions are accurate and internally consistent, I can only think of one criterion: Which description has more explanatory power? However, this may be the wrong question. I think one should rather embrace the fact there are many possibly equally good consistent but mutually exclusive descriptions of the real world and one should use whichever is better adapted to answer a particular problem. For example, a question regarding spacetime has most likely a simple answer in a covariant description, while a question regarding the observable algebra has most likely a simple answer in the Shape Dynamics description.

\section*{Acknowledgements} I want to thank J. Barbour for a careful proofreading of this contribution. Research at the Perimeter Institute is supported in part by the Government of Canada through NSERC and by the Province of Ontario through MEDT.

\section*{References}
\bibliography{koslowski-ae100prg}

\end{document}